\documentclass{PoS}

\newcommand{\half}{\frac{1}{2}}

\title{Spectrum of $4d$ ${\cal N}=1$ SYM on the lattice with light dynamical Wilson gluinos }

\ShortTitle{Spectrum of SYM on the lattice }

\author{K.~Demmouche, F.~Farchioni\speaker{}, A.~Ferling, G.~M\"unster, J.~Wuilloud\\
        
        University of M\"unster, Institute for Theoretical Physics  \\
        Wilhelm-Klemm-Strasse 9, D-48149 M\"unster, Germany\\
        E-mail: \email{k\_demm01@uni-muenster.de, farchion@uni-muenster.de}}

\author{I.~Montvay\\
           Deutsches Elektronen-Synchrotron DESY, Notkestr. 85, D-22603 Hamburg, Germany\\
          }
          
 \author{E.E.~Scholz\\
 Physics Department, Brookhaven National Laboratory, Upton, NY 11973, USA\\
 
  }

\abstract{
  We perform Monte Carlo investigations of the 4d ${\cal N}=1$ supersymmetric Yang-Mills (SYM)
  theory on the lattice with dynamical gluinos in the adjoint representation
  of the SU(2) gauge group.
  Our aim is to determine the mass spectrum of the low-lying bound states
  which is expected to be organised in supermultiplets in the infinite
  volume continuum limit.
  For this purpose we perform simulations on large lattices, up to an extension
  $L/r_0 \simeq 6$ where $r_0 \simeq 0.5\,\rm fm$ is the Sommer scale parameter.
  We apply improved lattice actions: tree-level improved Symanzik (tlSym) gauge
  action and in the later runs a Stout-smeared Wilson fermion action.
  The gauge configuration samples are prepared by the Two-Step Polynomial Hybrid
  Monte Carlo (TS-PHMC) update algorithm.
}

\FullConference{The XXVI International Symposium on Lattice Field Theory \\
        July 14 - 19, 2008\\
        Williamsburg, Virginia, USA}

\begin{document}

\section{Introduction}

The ${\cal N}=1$ supersymmetric Yang-Mills (SYM) theory is the minimal SUSY extension of the 
SU($N_c$) Yang-Mills theory. The fermionic degrees of freedom, the {\em gluinos}, are given by the superpartners 
of the gauge fields $A_\mu^a$ ({\em gluons}) and are described by Majorana spinors
$\lambda_a$ ($a=1\dots N_c^2-1$) transforming according to the adjoint representation of the gauge group.
SYM is characterized by a rich low-energy dynamics with interesting aspects as
confinement and spontaneous breaking of a discrete chiral symmetry  -- a continuum U(1) chiral symmetry is missing
at the quantum level due to the Adler-Bell-Jackiw anomaly. SYM can be related to QCD with a single quark flavour
($N_f=1$ QCD), where the Majorana spinor is replaced by the single Dirac spinor. The latter model is also object of
investigation by our collaboration \cite{nf1}.

This work represents a continuation of a long-standing project of the DESY-M\"unster-Roma Collaboration (DMRC)
for the simulation of SU(2) SYM,
see \cite{MontRev} for a review and \cite{fed-peetz-res} for more recent results. 
Following \cite{CV} we apply the Wilson approach, which has been proved to be successful 
in lattice QCD computations in spite of its known limitations.
SUSY is broken by the lattice discretisation and, in the Wilson approach, by the Wilson
term. It is expected to be recovered in the continuum limit by properly tuning the only relevant parameter, 
the bare gluino mass, to a critical value corresponding to massless gluinos.
Another (related) inconvenience of the Wilson discretisation, namely
a non positive-definite fermion measure even for positive gluino masses, turns out to have no appreciable 
impact in practical applications.

Past simulations of DMRC were performed on quite fine lattices,\footnote{We use QCD units 
for setting the scale. The Sommer scale parameter is fixed to the value $r_0\equiv 0.5$ fm.}
$a\simeq 0.08\,{\rm fm}$, but in a small volume, $L\simeq 1\,{\rm fm}$; 
this setup was appropriate for the study of the 
SUSY Ward identities \cite{susyWT}, also valid in a finite volume. 
We now concentrate on the mass spectrum of bound states for which
low-energy effective theories predict a reorganisation of the masses in two supermultiplets 
at the SUSY point \cite{VY,FaGaSc}. 
Thanks to a new more efficient simulation algorithm (see below) and enhanced 
computing resources we are now able to accumulate relevant statistics on larger lattices.
Our present series of numerical simulations are performed on $16^3\cdot32$ and $24^3\cdot48$ lattices 
with lattice spacing $a\simeq 0.125\,{\rm fm}$.
The lattice extension $L\simeq 2-3\,{\rm fm}$ is expected to be large enough to allow control over
finite volume effects on the bound states masses. Simulations on finer lattices are planned.

\section{Lattice formulation and algorithms}
\subsection{Lattice formulation}

We apply the tree-level Symanzik (tlSym) improved gauge action for the gauge part including
rectangular Wilson loops of perimeter six:
\begin{equation}
S_g=\beta (c_0 \sum_{pl}\lbrace  1-\frac{1}{N_c}\mbox{ReTr}U_{pl} \rbrace + c_1 \sum_{rec} \lbrace  
1-\frac{1}{N_c}\mbox{ReTr}U_{rec} \rbrace )\; ,
\end{equation}
with $c_0=1-8c_1$ and $c_1=-1/12$ in the case of tlSym action. 

The contribution of the gluino to the effective gauge 
action is given by
\begin{equation}
S^{\mbox{\it\scriptsize eff}}_{\tilde g}=-\frac{1}{2}\mbox{logdet}Q[U]  \;,   \label{CV action}
\end{equation}
where $Q$ is the non-hermitian Dirac-Wilson fermion matrix defined by
\begin{equation}
Q_{xy}^{ab}[U]=\delta_{xy}\delta^{ab}-\kappa\sum_{\mu=1}^4 (\delta_{x,y+\hat\mu}(1+\gamma_\mu)V_\mu^{ab}(y)+
\delta_{x+\hat\mu,y}(1-\gamma_\mu)V^{Tab}_\mu(x));
\label{fermion matrix}
\end{equation}
$V_\mu(x)$ is the adjoint gauge link, a real orthogonal matrix:
\begin{equation}
V_\mu^{ab}[U](x)=2\mbox{Tr}\lbrace U^\dagger_\mu(x)T^aU_\mu(x)T^b\rbrace=V^{*ab}_\mu(x)=[V^{-1ab}_\mu(x)]^T.
\end{equation}
($T^a$ are the generators of the SU($N_c$) group. In case of SU(2) 
one has $T^a=\frac{1}{2}\sigma^a$ with the Pauli matrices $\sigma^a$.)
Observe that ${\rm det} Q\geq 0$ for Majorana fermions.
In our recent simulations the links $U_{x,\mu}$ in the Wilson-Dirac
 operator Eq.~(\ref{fermion matrix}) are replaced by stout-smeared links \cite{MorningstarPeardon},
 which are defined as
\begin{equation}\label{stout}
U^{(1)}_{x,\mu} \equiv U_{x,\mu}\,\exp\left\{
\half\left( \Omega_{x,\mu} - \Omega^\dagger_{x,\mu} \right) -
\frac{1}{4}{\rm\,Tr}
\left( \Omega_{x,\mu} - \Omega^\dagger_{x,\mu} \right) \right\}\ ,\quad \Omega_{x,\mu} 
\equiv U^\dagger_{x,\mu} C_{x,\mu}
\end{equation}
 where
$C_{x,\mu}$ is a sum over staples
\begin{equation}
C_{x,\mu} \equiv \sum_{\nu\ne\mu} \rho_{\mu\nu} \left(
U^\dagger_{x+\hat{\mu},\nu} U_{x+\hat{\nu},\mu} U_{x,\nu} +
U_{x-\hat{\nu}+\hat{\mu},\nu} U_{x-\hat{\nu},\mu} U^\dagger_{x-\hat{\nu},\nu}
\right) \ .
\end{equation}
We apply one step of smearing with smearing parameters $\rho_{\mu\nu}=0.15$ in all lattice directions. 
%%%%%%%%%%%%%%%%%%%%%%%%%%%
   
 \subsection{Updating algorithm}

 The factor in front of $\log\det(Q)$ in the Curci-Veneziano action 
 reproduces the absolute value of the Pfaffian for Majorana fermions.
 Effectively, it corresponds to a flavour number $N_f=\half$.
 The sign of the Pfaffian can be included in a reweighting procedure (see below).
% Apart from the sign of the Pfaffian, one can replace $\det(Q)$ by
% $\{\det(Q^\dagger Q)\}^\half$.
 In our numerical simulations we use the two-step polynomial hybrid
 Monte Carlo (TS-PHMC) algorithm \cite{Montv} which is based
 on a two-step polynomial approximation \cite{Mont96}.
 The fermion determinant is represented  as
\begin{equation}
\mbox{det}(Q)^{N_f}= \lbrace\mbox{det}(Q^\dagger Q)\rbrace^{N_f/2} 
\simeq \frac{1}{\mbox{det }P_{n_1}(Q^\dagger Q)P_{n_2}(Q^\dagger Q)}
 \end{equation}
 with the condition on the polynomials $P_n$
\begin{equation}
\lim_{n_2\rightarrow\infty}P_{n_1}(x)P_{n_2}(x)=x^{-N_f/2}, \;\;\; x \in [\epsilon,\lambda],
\end{equation}
 where the interval covers the spectrum of $Q^\dagger Q$.
 The order of the first polynomial $n_1$ is chosen as low as possible
 provided that the acceptance in the accept-reject step done with $P_{n_2}$,
 after a sequence of PHMC trajectories prepared with $P_{n_1}$, is sufficiently high
 (in our cases 80-90\%).

 In earlier simulations of the DESY-M\"unster collaboration the two-step
 multi boson (TSMB) algorithm \cite{Mont96} was used for the SYM theory.
 There the updating is performed by heatbath and overrelaxation sweeps
 for the pseudofermions and Metropolis and heatbath sweeps for gauge field.
 In case of the TS-PHMC algorithm, which is more effective in producing
 short autocorrelations among the gauge configurations, PHMC trajectories
 are created by applying the Sexton-Weingarten integration scheme with
 multiple time scales and, as usual, Metropolis accept-reject at the end
 of each trajectory.
 After the update sweeps/trajectories the second high-precision polynomial
 $P_{n_2}$ is used in a stochastic {\em noisy correction step} with a
 global accept-reject condition.
 The polynomial $P_{n_2}$, which is an approximation for
 $[x^{-N_f/2}P_{n_1}]^{-1}$, has to be highly precise so that the error
 of the approximation is negligible compared to the overall statistical
 error, provided that the spectrum of $Q^\dagger Q$ is in the interval
 $[\epsilon,\lambda]$.

 If some of the eigenvalues is outside $[\epsilon,\lambda]$ then the
 approximations are not precise enough and a reweighting procedure has
 to be applied by an appropriately chosen high order polynomial.
 This gives a reweighting factor $C[U]$.
 In this reweighting procedure the sign of the {Pfaffian} of the fermion
 matrix can also be included in the measurement as follows:
\begin{equation}
\langle A \rangle=\frac{\langle {\rm sign}\mbox{Pf}[U]\, C[U]\, 
A[U]\rangle_g}{\langle {\rm sign}\mbox{Pf}[U]\, C[U] \rangle_g} \ .
\end{equation}
%

 %\section{Simulation setup and mass spectrum}
 \section{Simulation details}

The algorithmic parameters of our TS-PHMC runs are summarised in Table \ref{Table PHMC}. 
The runs are performed at a single lattice spacing corresponding to $\beta=1.6$.
The measured Sommer parameter for the lighter gluino masses is about $r_0/a\simeq 4$. 
By using the QCD scale as mentioned in the Introduction we obtain 
$a\simeq 0.125\,{\rm fm}$ and a box-size $L\simeq 2\,\mbox{fm}$ 
on the $16^3\cdot 32$ lattices and $L\simeq 3\,\mbox{fm}$  on the $24^3\cdot 48$ lattices. 
Comparison of results from ensembles $C_{(a,b)}$ on the $16^3\cdot32$ lattice
and $\bar C$ on the $24^3\cdot48$ lattice at the same gluino 
mass allows the study of finite size effects. 

An indication of the lightness of the gluino is given by the {\em adjoint pion} mass, $M_{a\mbox{-}\pi}$,
extracted from the connected part of correlator of the pseudoscalar gluino bilinear (see below). 
The lightest simulated adjoint pion mass (run $D$) corresponds to $M_{a\mbox{-}\pi}\simeq 353(20)$ MeV in QCD units. 

Compared with the previously used TSMB algorithm, TS-PHMC displays 
a substantially improved update efficiency with shorter plaquette 
autocorrelation times $\tau^{plaq}$.
This is particularly true in the light gluino regime where the efficiency 
of TSMB undergoes a strong depletion.

Some runs required the computation of the correction factor and the determination of the Pfaffian sign. 
We computed the correction factors for all runs in Table \ref{Table PHMC} with 
the exception of runs $A$, $B$, $A_s$ and $B_s$. 
Most signs of the Pfaffian are positive; the run with largest number of negative Pfaffians is
run $D$ where we found 15 configurations out of 5160 
with negative sign. In most cases the effect of the correction factor turned out to be negligible. 
This was not the case in run $D$ where the effect of $C[U]$ was important for the masses of adjoint 
meson bound states and the gluino-glueballs. All quoted statistical errors of the measured quantities 
were estimated by the $\Gamma$-method \cite{gamma method}. 

\begin{table}[t]
\caption{\em TS-PHMC runs algorithmic parameters with tlSym at $\beta=1.6$.
Runs labelled with $s$ have been performed with Stout-links;
$\delta\tau$ refers to the total trajectory length and $A_{NC}$ is the noisy correction acceptance.
}

\begin{center}

\renewcommand{\arraystretch}{1.0}

\begin{tabular}{l l|ccccccccc}
\hline\hline

Run    &$L^3.T$& $\kappa$ & \# Traj.     & $\delta\tau$ & $A_{NC}$ \% &  $\tau^{plaq}$ &$\epsilon$ & $\lambda$ & $n_1$ & $n_2$ \\ \hline
 $A$     & $16^3.32$   &  0.1800      &  2500    & 1.05 &   95.6 &   7.5 & $4.25\cdot 10^{-3}$ & 3.4 & 50 &100     \\
 $B$     & $16^3.32$   &  0.1900      &  2700    & 1.05 & 96.4 &   3.08 & $9.5\cdot 10^{-4}$   & 3.8 &  80 & 300 \\
 $Ca$   & $16^3.32$    &  0.2000    &  1973    & 0.99 & 82.9&   4.91   & $5.0\cdot 10^{-5}$ & 4.0 & 200 & 700 \\
 $Cb$   & $16^3.32$   &  0.2000    &  8874    & 0.99  & 88.3 & 27.6  & $5.0\cdot 10^{-5}$ & 4.0 & 200 & 700  \\
  $D$     & $16^3.32$  &  0.2020    &  6947    & 0.56 & 88.5&  45.7 & $1.0\cdot 10^{-6}$ & 4.0 & 800 & 2700  \\
 \hline\hline
$\bar A$ & $24^3.48$ &  0.1980   &  1480    & 0.9   & 89.6 & 4.64  & $1.0\cdot 10^{-4}$& 4.0 &  200 & 600           \\
$\bar B$ & $24^3.48$ &  0.1990   &  1400   &  0.9  & 88.7 &  2.65 & $4.0\cdot 10^{-5}$ & 4.0 & 270 & 800           \\
$\bar C$ & $24^3.48$ &  0.2000  &  6465    & 1.0  & 88.6  &  7.4  & $2.0\cdot 10^{-5}$ & 4.0 & 350 & 1000          \\
\hline\hline
% C3    & $32^3.64$   & 1.6 &  0.2000    &  in Run  &               &   \\
 $A_s$ & $24^3.48$   &  0.1500    &  370   &  1.0  &  97.3  & 3.5  & $5.5\cdot 10^{-5}$   & 2.2& 200 & 600  \\
 $B_s$ & $24^3.48$   &  0.1550    & 1730  &  1.0  & 95.6 &  8.04    &$5.5\cdot 10^{-5}$   & 2.2   & 200 & 600 \\
 $C_s$ & $24^3.48$  &  0.1570    &  2110  &  1.0 & 92.4 & 7.2  & $5.5\cdot 10^{-6}$ & 2.2 & 400 & 1200 \\

\hline\hline

\end{tabular}
\end{center}
\label{Table PHMC}
\end{table}

\section{Bound states}%

The bound states masses are computed from the zero-momentum 
correlation function of the corresponding interpolating operator ${\cal O}$. 

{\bf i) Adjoint mesons.}\ \
Low-energy theories \cite{VY,FaGaSc} predict a Wess-Zumino supermultiplet containing
colourless composite states of 
two gluinos. 
Such states have spin-parity quantum numbers $0^-$ and $0^+$. In analogy to flavour singlet QCD we denote the former 
$a$-$\eta^\prime$ and the latter $a$-$f_0$. To project these states on the lattice 
we used the gluino bilinear operators ${\cal O}=\bar\lambda\Gamma\lambda$ where $\Gamma=\gamma_5,1$ 
respectively. The resulting gluinoball propagator consists of connected and disconnected contributions:
\begin{equation}
C_{\Gamma}(t) = \frac{1}{V_s}\sum_{\vec x,\vec y} \left\langle \underbrace{\mbox{Tr}_{sc}[\Gamma Q_{xx}^{-1
}] \mbox{Tr}_{sc}[\Gamma Q_{yy}^{-1}]}_{\mbox{disconnected}}-2\underbrace{\mbox{Tr}_{sc}[\Gamma Q_
{xy}^{-1}\Gamma Q_{yx}^{-1}]}_{\mbox{connected}} \right\rangle 
- \frac{1}{V_s}\left\langle \frac{1}{T}\sum_t\sum_{\vec x} \mbox{Tr}_{sc} [\Gamma {Q_{xx}^{-1}}]\right\rangle^2\!\!.
\label{meson correlator}
\end{equation} 
The connected term can be used to extract the adjoint pion mass $M_{a\mbox{-}\pi}$
(the last term in Eq.~(\ref{meson correlator}) vanishes for $\Gamma=\gamma_5$).
OZI arguments \cite{CV} suggest   $M^2_{a\mbox{-}\pi}\propto m_{\tilde g}$ for light gluinos.

The disconnected propagators were computed by using the Stochastic Estimators Technique (SET) 
in the spin dilution variant to reduce the large variance. As it is the case in QCD, 
the disconnected diagrams are intrinsically noisier than the connected ones
and dominate the level of noise in the total correlator.
We performed tests on few configurations with up to $N=40$ noisy estimators 
in order to study the limit at which only the gauge noise dominates the statistical 
error of the disconnected correlator. 
The optimal number of estimators was finally fixed between 16 and 22 estimates for all runs. 
In the case of $a$-$\eta^\prime$ reasonable signal-to-noise ratio is obtained allowing 
the extraction of the mass from the mass fit. This was not possible for the $a$-$f_0$;
the latter has a nonzero overlap with the vacuum, and hence its correlator has a 
non-vanishing vacuum contribution
(the last term in Eq.~(\ref{meson correlator})). This results in a much worse
signal-to-noise ratio, and the effective mass could only be 
extracted at very short time separations.
In the future we shall consider new variance reduction techniques 
for a more precise computation of the disconnected diagrams.

{\bf ii) Gluino-glueballs.}\ \ 
The gluino-glueballs ($\tilde g\mbox{-}g$) are spin-$\frac{1}{2}$ colour 
singlet states of a gluon and a gluino. They are supposed to complete the Wess-Zumino 
supermultiplet of the adjoint mesons \cite{VY}.
The full correlator is built up from plaquettes connected
by a gluino propagator line: 
\begin{equation}
C^{\alpha\beta}_{\tilde g\mbox{-}g}(\Delta t)=-\frac{1}{4}\sum_{\vec x, \vec y  }\sum_{i,j,k,l}\left\langle  
\sigma_{ij}^{\alpha\alpha^\prime}\mbox{Tr}[U_{ij}(x)\sigma^a]Q^{-1}_{xa\alpha^\prime,yb\beta^\prime}
\mbox{Tr}[U_{kl}(y)\sigma^b]\sigma_{kl}^{\beta^\prime\beta}\right\rangle. 
\end{equation} \label{gluino-glue correlator}
The above correlator is a matrix with two independent components in Dirac space:
\begin{equation}
 C^{\alpha\beta}_{\tilde g\mbox{-}g}(\Delta t)=C_{1}(\Delta t)\delta^{\alpha\beta}+C_{\gamma_4}(\Delta t)\gamma_4^{\alpha\beta}\;.
\end{equation}
We see agreement in the masses extracted from each component, 
and we choose the time antisymmetric component $C_{1}$ to fit the masses.   
We apply APE smearing for the links and Jacobi smearing for the fermion fields 
in order to optimise the signal-to-noise ratio and to obtain an earlier plateau 
in the effective mass.

{\bf iii) Glueballs.}\ \ 
According to \cite{FaGaSc},  $0^{\pm}$ glueballs are expected to be members of a second Wess-Zumino
supermultiplet. Their study in SYM
presents difficulties which closely resemble those encountered 
in glueball spectroscopy in QCD and, fortunately, can be overcome with the same type of techniques, 
namely, APE smearing with the variational method. Also in this case we use the simplest 
interpolating operator for the scalar glueball $0^{++}$ built from single space-like plaquette. 

It turns out, however, that the present statistics is not enough
to obtain a reliable determination of the glueball masses. Therefore, we are planning to increase
the statistics.

\begin{figure}[t]
\begin{center}
\includegraphics[angle=0,width=.7\linewidth]{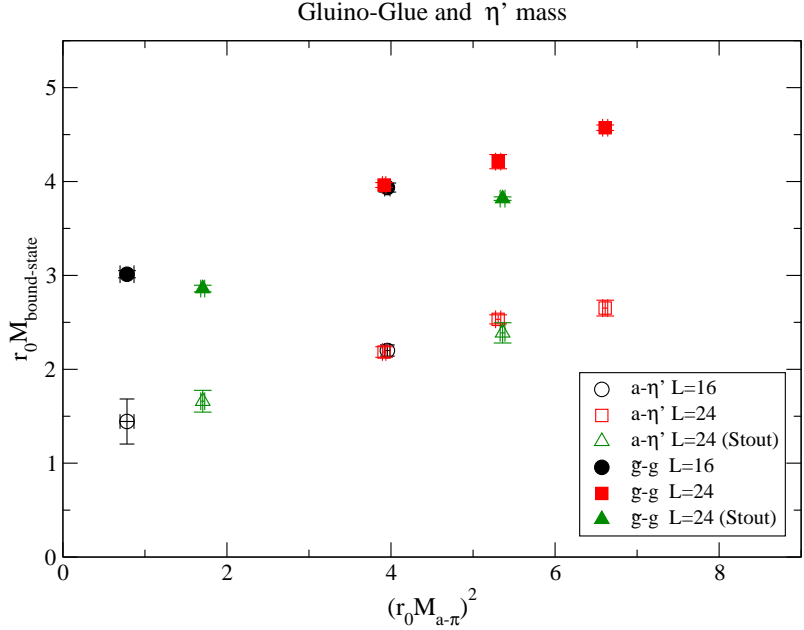}
\caption{The masses of the low lying bound states of the ${\cal N}=1$ SU(2) SYM theory. 
Shown are the masses multiplied by $r_0$ as a function the squared adjoint pion mass. 
}
\label{fig:spectrum}
\end{center}
\end{figure}

{\bf Results.}\ \ 
The masses of the $a$-$\eta^\prime$ and the gluino-glueball are displayed in 
Fig.~\ref{fig:spectrum} as a function of the squared adjoint pion mass in units of the Sommer scale parameter.
Both bound state masses appear to be characterized by a linear dependence on 
$(r_0M_\pi)^2$. The  gluino-glueball turns out to be appreciably heavier  (50\%)
than the $a$-$\eta^\prime$.
Runs with and without Stout-smearing  give consistent results for the $a$-$\eta^\prime$ mass, while a
discrepancy is observed for the gluino-glueball, which can be interpreted as an $O(a)$ 
discretisation effect. The comparison of the two runs at $(r_0M_\pi)^2\simeq 4$ in 
$(2\,\mbox{fm})^3$ and $(3\,\mbox{fm})^3$ volumes reveals small finite volume effects.
The linear extrapolation of the  $a$-$\eta^\prime$ mass to massless adjoint pion 
(including the four lightest points) gives: $r_0M_{a\mbox{-}\eta^\prime} = 1.247(48)$
[$499(20)\,\mbox{MeV}$]. A rough estimate for the gluino-glueball  
gives  $r_0M_{\tilde g\mbox{-}g} = 2.5-3$ [$1000-1200\,\mbox{MeV}$].
\vspace{1mm}

\section{Summary and conclusions}

New results on the spectrum of bound states have been obtained by means of the TS-PHMC
algorithm with noisy correction.
The new algorithm, with an improved gauge action and Stout-smeared links,
allows to obtain a significantly better performance compared to the 
previously used TSMB algorithm.
This allows us to simulate 
in significantly larger volumes of $(2\,\mbox{fm})^3$ and $(3\,\mbox{fm})^3$.
Our first results for the masses are within errors equal
in these two cases implying that these volumes are large enough for
the study of the particle spectrum.
We shall systematically investigate the finite volume effects
in future publications.

According to the low-energy effective theory of \cite{VY} the $a$-$\eta^\prime$ and the gluino-glueball
belong to the same supermultiplet and therefore should be degenerate in the SUSY limit.
However our preliminary results show a gluino-glueball mass systematically heavier than the $a$-$\eta^\prime$
until the lightest simulated gluino mass in the weakly broken SUSY region. 
Whether this outcome is a discretisation artifact or a physical effect will
become clear in future studies at finer lattice spacings. If the latter case applies,
the interpolating gluino-glueball operator could have dominant overlap with a member 
of a higher supermultiplet.
The complete identification of possible supermultiplets requires the
inclusion of  the masses of the glueballs and the scalar $a$-$f_0$ bound states.

\vspace{1mm}

The computations were carried out on Blue Gene L/P and JuMP systems at JSC J\"ulich, Opteron PC-cluster 
at RWTH Aachen and the ZIV PC-cluster of the university of M\"unster (Germany).

\end{document}